\documentclass[12pt]{article}
\usepackage{graphicx} % Required for inserting images
\usepackage{subcaption}
\usepackage{cite}
\usepackage{amsmath} 
\usepackage[utf8]{inputenc}
\usepackage[a4paper, margin=1in]{geometry}
\usepackage[colorlinks=true, linkcolor=blue, citecolor=blue, urlcolor=blue]{hyperref}

\usepackage{caption}
\begin{document}
\title{Ising Model with Power Law Resetting}

\author{
Anagha V K$^{1}$ and Apoorva Nagar$^{1}$\\[6pt]
\small $^{1}$Department of Physics, Indian Institute of Space Science and Technology,\\
\small Thiruvananthapuram 695547, India
}

\date{\today}

\maketitle

\begin{abstract}
We investigate the nonequilibrium dynamics of the nearest-neighbour Ising model subjected to stochastic resetting, where the system is intermittently returned to an initial configuration with magnetisation $m_0$, with the inter-reset times drawn from the power law distribution $\alpha \tau_0^\alpha / \tau^{\alpha+1}$. The heavy-tailed resets generate magnetisation distributions that differ significantly from both equilibrium dynamics and the previously studied Ising model with exponentially distributed reset times. In two dimensions, for $T > T_C$, we find a quasi-ferro state for all $\alpha$, marked by a double-peaked distribution that diverges at $m=0$ and $m=m_0$; no steady state exists for $\alpha < 1$, while a stationary state emerges for $\alpha > 1$. For $T < T_C$, power law resetting produces two distinct regimes separated by a crossover exponent $\alpha^* = 1-c$: a single-peak ferromagnetic phase localised at $m_{eq}$ for $\alpha < \alpha^*$, and a dual-peak ferromagnetic phase with divergences at $m_{eq}$ and $m_0$ for $\alpha > \alpha^*$. Analytic results in one and two dimensions, supported by simulations, yield a rich phase diagram in the $(T,\alpha)$ plane and reveal how heavy-tailed resetting generates nonequilibrium phases very different from those seen in the case of exponential resetting.
\end{abstract}

\section{Introduction}
\label{sec:1}
Stochastic resetting is a process in which an evolving dynamical system is suddenly brought back to a pre-decided state at random times and allowed to evolve via its natural dynamics in between two such resets. This seemingly simple mechanism is found to have intriguing effects in system dynamics, which often lead to non-trivial steady state behaviour. This concept has emerged as a versatile tool for modelling and analysing different phenomena since it was originally introduced in the context of a diffusion search problem \cite{evans2011diffusion}. In that particular work, it was shown that this mechanism leads to a more efficient searching process, and the first passage time can be minimised by tuning the resetting rate. It was also found that this process breaks the detailed balance and leads to the emergence of a non-equilibrium steady state. Since this pioneering work, stochastic resetting has attracted significant research interest. Early works examined diffusion with space-dependent or randomly distributed resets ~\cite{evans2020stochastic, evans2011diffusion1,evans2014diffusion} and the optimisation of first-passage times, including scenarios with partial absorption \cite{whitehouse2013effect,evans2013optimal}. Subsequent studies extended the concept to discrete-time jumps \cite{kusmierz2014first}, memory-dependent resetting \cite{boyer2017long}, fractional Brownian motion \cite{majumdar2018spectral}, and resets to extremal positions \cite{majumdar2015random}. Non-instantaneous or finite-speed resets have also been considered to account for spatio-temporal constraints \cite{pal2019time, pal2019invariants, gupta2020stochastic, gupta2021resetting}, along with complex geometries such as bounded domains \cite{pal2019first, bonomo2021first, chatterjee2018diffusion}, comb structures \cite{singh2021backbone}, and Sisyphus random walks \cite{montero2016directed}. Beyond classical diffusion, stochastic resetting has been explored in quantum dynamics \cite{mukherjee2018quantum}, run-and-tumble particles \cite{evans2018run}, and telegrapher’s processes \cite{masoliver2019telegraphic}. Experimental confirmations for diffusion with resetting have been reported using optically trapped colloidal particles \cite{besga2020optimal, tal2020experimental}. Applications extend across disciplines: in biology for transcriptional pauses, protein–DNA target search, and active intracellular transport \cite{roldan2016stochastic,cherstvy2008protein,bressloff2020modeling}, in ecology for animal relocation and movement patterns \cite{boyer2014random, kenkre2021theory}, in computer science for randomised search algorithms \cite{ginsberg1993dynamic,montanari2002optimizing}, in psychology for visual search and pattern recognition \cite{noton1971scanpaths,eckstein2011visual}, and in finance for reset options and warrant valuations \cite{cheng2000analytics,gray1997valuing}. The wide-ranging effects of stochastic resetting on search efficiency, first-passage behaviour, and nonequilibrium dynamics make it a rich and rapidly expanding area of research.
\par
While most studies of stochastic resetting have focused on non-interacting systems, its effects in interacting systems are particularly important, as they reflect more realistic dynamics \cite{nagar2023stochastic}. To understand the range of behaviours induced by resetting, researchers have explored diverse interacting systems, from simple two-particle setups to complex many-particle models. In the simplest case of two-particle systems, stochastic resetting has been investigated in predator-prey models \cite{evans2022exactly,mercado2018lotka}, interacting Brownian motion \cite{falcao2017interacting}, and diffusion with birth-death dynamics \cite{da2018interplay,da2019diffusions,da2021diffusion}. Studies on many-particle systems have examined how stochastic resetting affects models such as the symmetric simple exclusion process (SSEP) \cite{basu2019symmetric,sadekar2020zero}, the totally asymmetric simple exclusion process (TASEP) \cite{karthika2020totally}, fluctuating interfaces \cite{gupta2014fluctuating,gupta2016resetting}, coagulation-diffusion processes \cite{durang2014statistical}, zero-range processes \cite{grange2020non}, and the Ising model \cite{magoni2020ising}. Further, in such systems, it is possible to have dynamics involving the reset of a part of the system instead of the whole system. The effects of such local resetting have also been explored in binary aggregation \cite{grange2021aggregation}, SSEP \cite{miron2021diffusion}, and TASEP \cite{pelizzola2021simple}. In this work, we focus on the nearest-neighbour Ising model with stochastic resetting, which is an archetypal model to study phase transition, making it an ideal platform to investigate how resetting modifies dynamics and steady-state behaviour.
\par
In a previous work, Magoni et al. \cite{magoni2020ising} studied the Ising model undergoing stochastic resetting at a constant rate. They considered a nearest-neighbour Ising model, which evolves according to Glauber dynamics in the absence of reset. The system, which is trying to evolve to its equilibrium state, is repeatedly brought back to a configuration with magnetisation $m_0$ after random intervals of time which are distributed exponentially ($re^{-r\tau }$). The quantity under observation was the distribution of the magnetisation of the system at long times. They found that the introduction of resetting leads to a nontrivial phase behaviour. The system can exist in a `ferro' phase $(T<T_c)$, where the stationary magnetisation distribution $P_{\text{stat}}(m)$ develops a clear gap at $m=0$ and peaks at a finite value of $m$; or a `para' phase $(T>T_c,\, r(T)<r^{*}(T))$ where $P_{\text{stat}}(m)$ instead exhibits a divergent peak at $m=0$ with no gap. Apart from these two phases, which are broadly similar to the usual ferromagnetic and paramagnetic phases, their work also shows the emergence of a `pseudoferro' phase for $T>T_c$ and $r(T)>r^{*}(T)$, where $P_{\text{stat}}(m)$ remains gapless but now vanishes at $m=0$ and peaks at a finite value of $m$ ($m=m_0$). The introduction of resetting thus leads to a richer and more interesting behaviour than the usual Ising dynamics.
\par
In this work, we study the Ising model with stochastic resetting such that the time intervals between the resets are distributed according to a power law $(\alpha \tau_0^{\alpha}⁄\tau^{\alpha+1})$. The paradigmatic nature of the Ising model makes it an ideal system to study the effects of novel dynamical features, while the choice of power law distributed time intervals is guided by their ubiquity in natural as well as human-made systems. Such waiting time distributions commonly appear in complex systems and lead to events that occur in bursts with long pauses in between. This type of behaviour is observed in diverse systems like earthquakes \cite{bak2002unified}, neuron firing \cite{kemuriyama2010power}, stock market fluctuations \cite{sornette2017stock,gontis2007modeling,lillo2003power}, and human activity patterns \cite{gandica2017stationarity,lee2021self,yu2023understanding}. Power law resetting introduces heavy-tailed intervals, allowing rare but arbitrarily long times between resets, which leads to a richer interplay between the system’s intrinsic relaxation and the resetting mechanism. Previous studies have shown that power law resetting can significantly alter system dynamics compared to exponential resets. In the context of diffusion, power law resetting produces a wide range of long-time behaviours: depending on the tail exponent $\alpha$, the motion may remain unbounded or become stationary \cite{nagar2016diffusion}, whereas exponential resetting always drives the system to a stationary state \cite{evans2011diffusion}. Similarly, in TASEP, exponential resetting leads to a steady state with monotonic density evolution, while power law resetting can generate long-lived, time-dependent scaling and non-monotonic density profiles when the tail exponent is small \cite{karthika2020totally}. These examples highlight how the choice of reset distribution qualitatively affects long-time dynamics, thus motivating this study.
\par
Our framework is the same as in \cite{magoni2020ising}. We consider a nearest neighbour Ising model with initial magnetisation $m_0$. The system evolves according to Glauber dynamics \cite{glauber1963time} towards its equilibrium state with magnetisation $m_{eq}$. In the absence of resetting, for dimensions (D) greater than one, the system reaches a ferromagnetic phase $(m_{eq}>0)$ for $T<T_C$ and a paramagnetic phase $(m_{eq}=0)$ for $T \geq T_C$ at long times. The distribution of magnetisation in equilibrium is a delta function centred at $m_{eq}$ corresponding to the temperature chosen. Now we introduce a resetting protocol. The system is brought back to a configuration with magnetisation $m_0$ (initial magnetisation) at random intervals of time which are distributed according to a power law, $\alpha \tau_0^\alpha⁄\tau^{\alpha+1}$. In between the resets, the system evolves via Glauber dynamics. This resetting process can be physically realised by a ‘rapid quench’ protocol as described in \cite{magoni2020ising} by replacing the exponential time interval distribution with a power law distribution. We observe the magnetisation $m(t) =\frac{1}{N} \sum_i \langle s_i(t) \rangle$  at time $t$ and analyse how it is distributed.
\par
\begin{figure}[ht]
    \centering
    \includegraphics[width=0.9\linewidth]{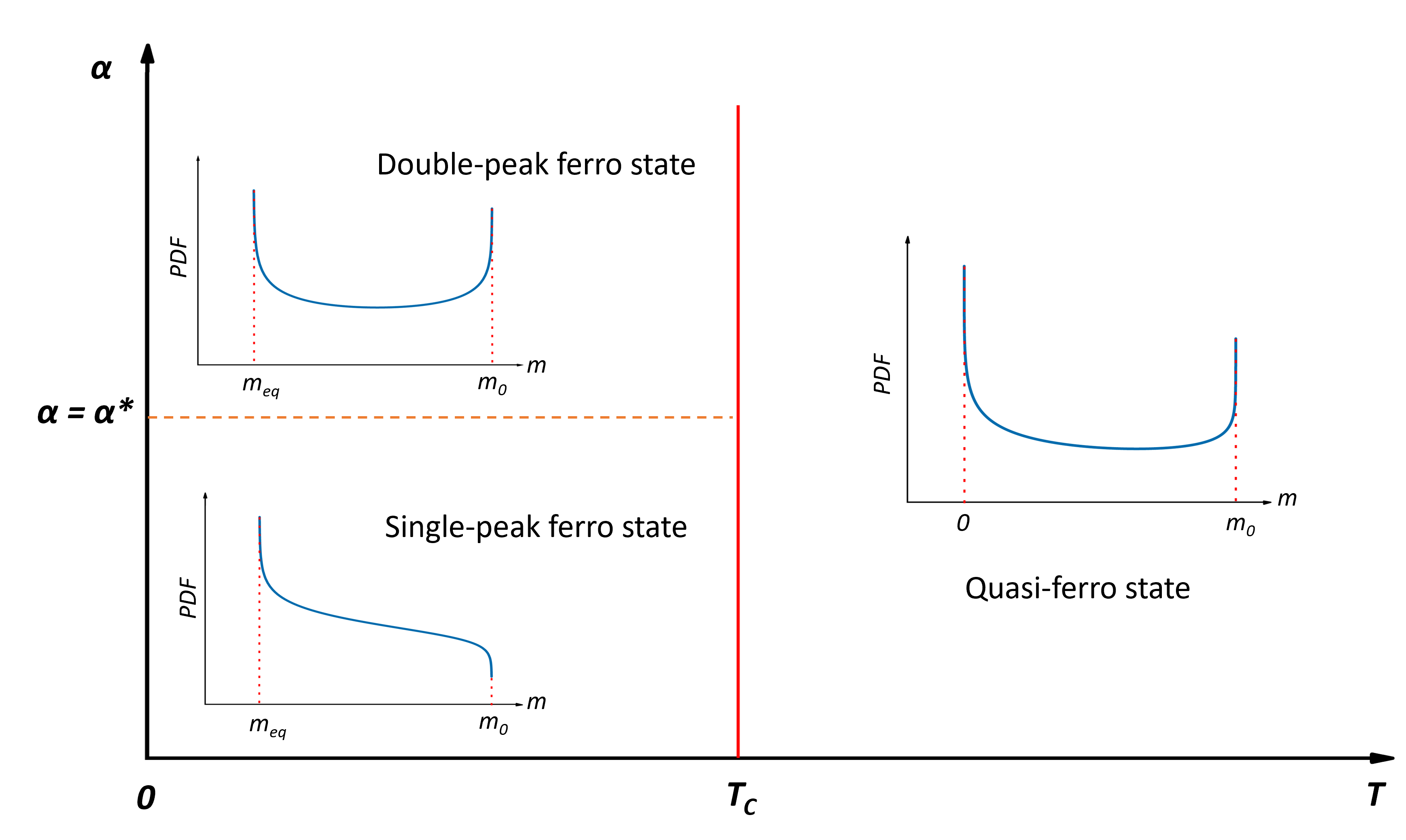}
    \caption{Phase diagram in the $(T,\alpha)$ plane for the 2D Ising model with power law stochastic resetting. 
        The schematic shows the qualitative forms of the magnetisation distribution in the different regimes. For $T>T_c$, the system exhibits a quasi-ferromagnetic state for all $\alpha$, characterised by a double-peaked distribution with divergences at $m=0$ and $m=m_0$. 
        For $T<T_c$, power law resetting gives rise to two distinct ferromagnetic phases separated by the 
        crossover line $\alpha=\alpha^*$ (dashed orange), where $\alpha^*=1-c$ and $c$ is the Glauber 
        dynamical exponent. For $\alpha<\alpha^*$, the distribution is single-peaked, diverging 
        only at the equilibrium magnetisation $m_{eq}$. For $\alpha>\alpha^*$, a double-peaked 
        ferromagnetic state emerges, with divergences at both $m_{eq}$ and $m_0$.}
    \label{fig:a}
\end{figure}

In our work, we see that the resetting at power law times leads to non-trivial magnetisation distributions which are distinctly different from those seen in the Ising model with exponentially distributed reset times. In 2D, the Ising model without reset exists in the paramagnetic ($T>T_C$) or ferromagnetic ($T<T_C$) phase. The introduction of power law resetting leads to the following behaviour: For $T>T_C$, we see a ‘quasi-ferro’ state (QF) in which for all values of $\alpha$, the probability distribution takes a double-peaked structure. For $\alpha<1$, the system does not reach a steady state. The probability density function at large times diverges at both the origin ($m=0$) and $m=m_0$. On the other hand, for $\alpha>1$, the system goes to a stationary state at long times, and the probability function continues to have the double-peak structure with divergences at $m=0$ and  $m=m_0$ as before. We thus see that the behaviour is very different from the paramagnetic state of the Ising model, as well as the `pseudoferro' phase seen in \cite{magoni2020ising}. For $T<T_C$, we see that depending on the value of $\alpha$, there is an emergence of two different phases. For $\alpha<1$, again the system does not go to a steady state in the large time limit. But for  $\alpha<\alpha^*$, we see a single-peaked magnetisation distribution at large times. The probability peaks only at $m=m_{eq}$  in this ‘single-peak ferro state’ (SPF). On the other hand, for  $\alpha>\alpha^*$, we again see a double-peaked probability function that diverges at $m=m_{eq}$ as well as at $m=m_0$  at long times. For $\alpha>1$, the system possesses a steady state with the probability function again peaking at $m_0$ and $m_{eq}$. So, in effect, for $\alpha>\alpha^*$, we have a ‘double-peak ferro state’ (DPF) which is different from the case of $\alpha<\alpha^*$. Figure~\ref{fig:a} shows a plot on the $(T,\alpha)$ plane indicating the various phases along with the behaviour of magnetisation distribution in each case. For $T<T_C$, we see a crossover line at $\alpha=\alpha^*$ (where $\alpha^*=1-c, 0<c<1$, where $c$ is a dynamical parameter corresponding to Ising-Glauber dynamics for $T<T_C$) that separates the two distinct phases we discussed above. For the 1D Ising model, the original Ising system exists in a paramagnetic state at all finite temperatures. Correspondingly, the introduction of the power law resetting leads to a QF state with a time-independent steady state distribution existing only for $\alpha>1$.  We establish these results using analytical solutions for the cases of 1D and 2D Ising models and validating them with numerical simulations.
\par
The paper is organised as follows: In section~\ref{sec:2}, we introduce the Ising model with power law resetting. We begin by briefly recalling the behaviour of the model without resetting and then present the renewal-based analytical framework used to incorporate stochastic resetting. Section~\ref{sec:3} focuses on the one-dimensional Ising model, where we derive the magnetisation distribution under resetting and compare our analytical predictions with numerical simulations. In section~\ref{sec:4}, we extend the analysis to the two-dimensional Ising model and examine how resetting affects the magnetisation distribution across different temperature regimes, again validating our results through simulations. Finally, we summarise our findings and outline possible directions for future work in section~\ref{sec:5}.

\section{Ising model with power law resetting}
\label{sec:2}
We consider a $d$-dimensional nearest-neighbour Ising model with $N$ sites and periodic boundary conditions. The Hamiltonian is given by $H=-J\sum_{\langle i,j \rangle} s_i s_j$. The initial state is chosen such that the magnetisation of the configuration equals $m_0$. For this, we generate a configuration by independently choosing the spins to take values $\pm 1$ with probability $(1\pm m_0)/2$, respectively (where $m_0\in [0,1]$). This is also the state to which the system resets repeatedly. In the absence of reset, the system follows Glauber dynamics, which consists of flipping a single spin with rate $w(s_i\to -s_i)=\frac{1}{1+e^{\beta \Delta E}}$,
where $\beta=(1/k_B T)$ and $\Delta E=2Js_i \sum_{j\in n.n.}s_j$ is the change in energy when the spin $s_i$ is flipped \cite{glauber1963time}. In one dimension, this rate takes the form $w(s_i\to -s_i)=(1/2)(1-\gamma s_i(s_{i-1}+s_{i+1})/2)$ where $\gamma = tanh(2\beta J)$ \cite{glauber1963time}.
\par
Now we introduce a resetting protocol in which the system is returned, at random times, to the initial state of magnetisation $m_0$, where the inter-reset times are drawn from a power law distribution: 
\begin{equation}
    \rho(\tau)=\frac{\alpha \tau_0^\alpha}{\tau^{\alpha+1}};\tau \in[\tau_0,\infty),\alpha>0
    \label{eq:1}
\end{equation}

In between resets, the system relaxes toward its equilibrium state corresponding to the temperature $T$ via Glauber dynamics. Our aim is to determine the probability distribution of the magnetisation, $m(t) =\frac{1}{N} \sum_i \langle s_i(t) \rangle$, at the observation time $t$. To do this, we employ a renewal approach: each reset restarts the dynamics, so the magnetisation distribution at time $t$ is solely determined by the evolution that follows the last reset prior to $t$. The corresponding renewal equation is
\begin{equation}
    P_r(m,t)=\int_0^t d\tau f(t,t-\tau)P_0(m,\tau)
    \label{eq:2}
\end{equation}
where $f(t,t-\tau)$ is the probability density at time $t>0$ for the last reset to have happened at the time instant $t-\tau; \tau \leq t$. $P_0(m,\tau)$ is the probability distribution of the magnetisation in the absence of resetting. The spatially averaged quantity $m(t)$ has a deterministic evolution and therefore this distribution is simply given by, $P_0(m,\tau)=\delta(m-m(\tau))$. For power law resets, the behaviour of $f(t,t-\tau)$ depends on the exponent $\alpha$ \cite{nagar2016diffusion};
For $\alpha <1$,
\begin{equation}
    f_{\alpha<1}(t,t-\tau)=\frac{sin(\pi \alpha)}{\pi} \tau^{-\alpha}(t-\tau)^{\alpha-1}
    \label{eq:3}
\end{equation}
and thus,
\begin{equation}
    P_r(m,t)=\frac{sin(\pi \alpha)}{\pi} \int_0^t d\tau \tau^{-\alpha}(t-\tau)^{\alpha-1}P_0(m,\tau)
    \label{eq:4}
\end{equation}
When $\alpha>1$, for $\tau \ge \tau_0$, we have
\begin{equation}
    f_{\alpha>1,\tau \ge \tau_0}(t,t-\tau)=\frac{1}{\tau_0} \left ( \frac{\alpha-1}{\alpha} \right)\left (\frac{\tau}{\tau_0}\right)^{-\alpha}
    \label{eq:5}
\end{equation}
In addition, the probability density of resetting in the small interval $[0,\tau_0]$ follows from normalisation of $f_{\alpha>1}(t,t-\tau)$;
$\int_0^{\tau_0}d\tau f_{\alpha>1,\tau < \tau_0}(t,t-\tau)=1-\int_{\tau_0}^{t}d\tau f_{\alpha>1,\tau \ge \tau_0}(t,t-\tau) $. By combining these contributions, we obtain
\begin{equation}
    P_r(m,t)=\frac{1}{\tau_0}\left (\frac{\alpha-1}{\alpha} \right) \int_{\tau_0}^t d\tau \left (\frac{\tau}{\tau_0}\right)^{-\alpha}P_0(m,\tau) + \left[ 1-\frac{1}{\alpha}\left(1-\left(\frac{\tau_0}{t}\right)^{\alpha-1}\right)\right]P_0(m,\tau_0)
    \label{eq:6}
\end{equation}

\section{Resetting in the 1D Ising model}
\label{sec:3}

For the 1D Ising model $T_C=0$, and the system shows a paramagnetic behaviour at all finite temperatures. We consider a one-dimensional lattice with periodic boundary conditions. In the absence of resets, the magnetisation $m(t)$ evolves deterministically via Glauber dynamics \cite{glauber1963time} as
\begin{equation}
    m(t)=m_0 e^{-(1-\gamma)t} , \gamma=tanh(2\beta J)
    \label{eq:7}
\end{equation}
Using this relaxation trajectory, the effect of stochastic resetting on the probability distribution of the magnetisation can be calculated using Eqs. ~\eqref{eq:4} (for $\alpha <1$) and  ~\eqref{eq:6} (for $ \alpha>1$). For $\alpha<1$, we have
\begin{equation}
    P_r(m,t)=\frac{sin(\pi \alpha)}{\pi} \frac{1}{m}\left[ ln\left(\frac{m_0}{m}\right)\right]^{-\alpha}\left[t(1-\gamma)-ln\left(\frac{m_0}{m}\right)\right]^{\alpha-1}
    \label{eq:8}
\end{equation}
We can see that the system does not reach a stationary state. The distribution at any finite time exhibits a characteristic double-peaked structure with divergences at both $m=0$ and $m=m_0$. As $m \to 0$, the divergence of the power law dominates over the decaying logarithm, and hence we get a peak at $m=0$. At $m=m_0$, we again have a divergent behaviour via the term $(ln \frac{m_0}{m})^{-\alpha}$.
\par
When $\alpha>1$, using Eqs.~\eqref{eq:7} and ~\eqref{eq:6}, we get
\begin{equation}
\begin{split}
     P_r(m,t)= \frac{1}{\left[\tau_0(1-\gamma)\right]^{1-\alpha}}\left (\frac{\alpha-1}{\alpha} \right) \frac{1}{m}\left[ ln\left(\frac{m_0}{m}\right)\right]^{-\alpha}
     \\+\left[ 1-\frac{1}{\alpha}\left(1-\left(\frac{\tau_0}{t}\right)^{\alpha-1}\right)\right] \delta(m-m_{\tau_0})
     \label{eq:9}
\end{split}
\end{equation}
At large times, the system goes to a stationary state with the distribution 
\begin{equation}
    P_r^{stat}(m)=\left (\frac{\alpha-1}{\alpha} \right) \left( \frac{1}{\left[\tau_0(1-\gamma)\right]^{1-\alpha}}\frac{1}{m}\left[ ln\left(\frac{m_0}{m}\right)\right]^{-\alpha}+\delta(m-m_{\tau_0})\right)
    \label{eq:10}
\end{equation}
Here also we see a divergence at $m=0$ and $m=m_0$. For both the cases, $\alpha<1$ and $\alpha>1$, we get double-peaked structures that diverge at $m=0 $ and $m=m_0$. Physically, this reflects the competition between two processes: the short bursts of resetting favour the initial state $m_0$, while the long tail of the power law supports the relaxation towards zero magnetisation.
\par
We can calculate the average of this magnetisation distribution as $\langle m\rangle=\int mP_r(m)dm$.
For $\alpha<1$, at a fixed time $t$,
\begin{equation}
\langle m\rangle=m_0 {}_1F_1\left(1-\alpha;1,-(1-\gamma)t\right)
\label{eq:a1}
\end{equation}
where ${}_1F_1\left(1-\alpha;1,-(1-\gamma)t\right)$ is Kummer's (confluent hypergeometric) function of the first kind.
For $\alpha>1$, 
\begin{equation}
\langle m\rangle=\left(\frac{\alpha-1}{\alpha}\right)\left(\left[\tau_0(1-\gamma)\right]^{\alpha-1}m_0 \Gamma(1-\alpha,\varepsilon)+m_{\tau_0}\right)
\label{eq:a2}
\end{equation}
where $\Gamma(1-\alpha,\varepsilon)$ is upper incomplete gamma function and $\varepsilon=(1-\gamma)\tau_0$. So we see that there exists an average non-zero magnetisation for any $m_0\ne 0$. We therefore call this phase a `quasi-ferro' state (QF), where we have a divergence at $m=0$ as expected for a paramagnetic state, but the other divergence at $m=m_0$ leads to finite magnetisation in the system for all $T>0$.
\par
\begin{figure}[htbp]
  \centering
  \begin{subfigure}[b]{0.5\textwidth}
    \includegraphics[width=\textwidth]{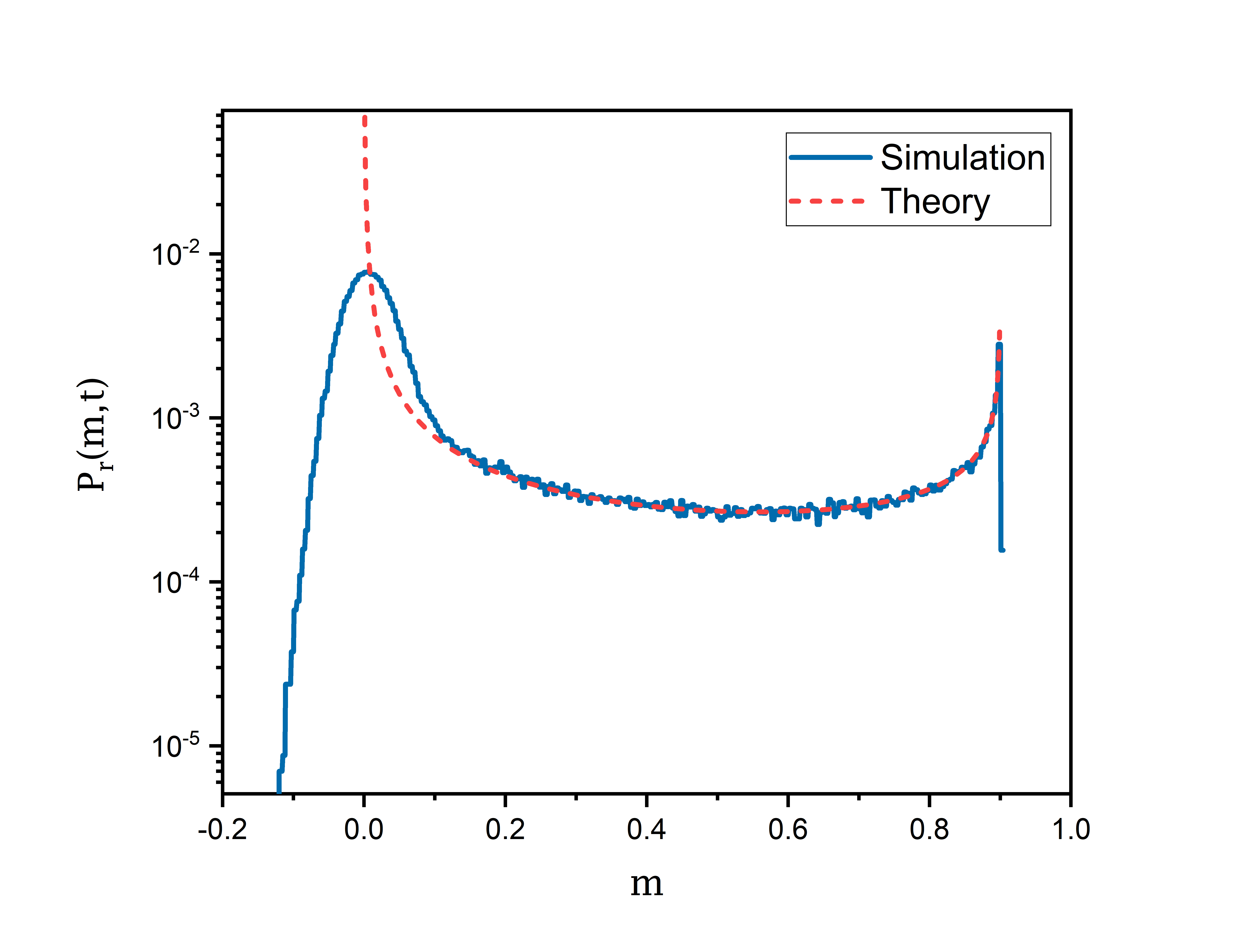}
    \caption{$\alpha<1$}
    \label{fig:1a}
  \end{subfigure}
  \hspace{-0.5cm}
  \begin{subfigure}[b]{0.5\textwidth}
    \includegraphics[width=\textwidth]{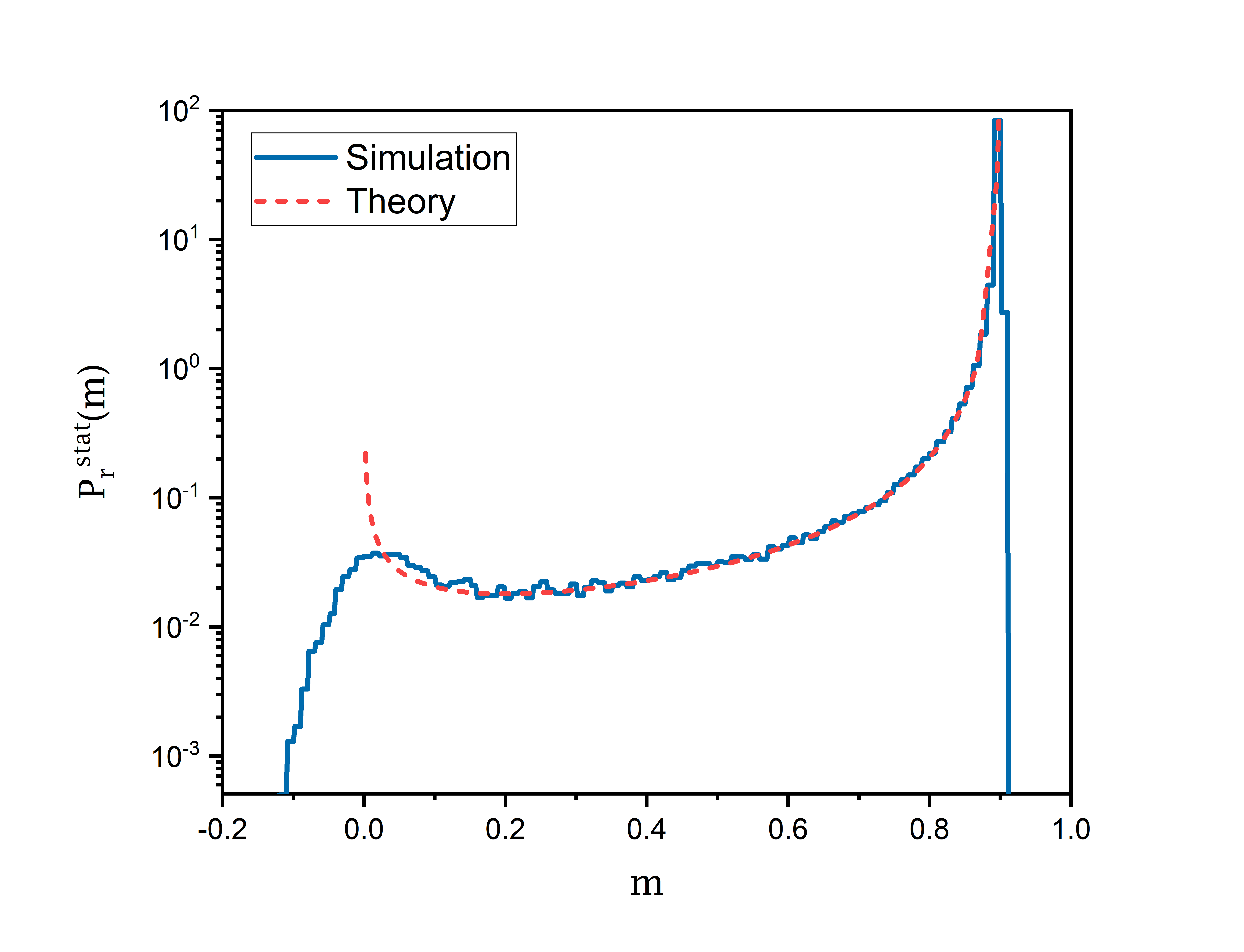}
    \caption{$\alpha>1$}
    \label{fig:1b}
  \end{subfigure}
  \caption{Plot of the PDF of the magnetisation in the 1D Ising model with power law resetting for different values of $\alpha$: (a) Finite-time distribution at $t=12$ for $\alpha=0.5$ ($\alpha<1$), lattice size $L=2000$, $m_0=0.9$ and $\gamma=0.6$, averaged over $10^5$ realisations. (b) Steady-state distribution for $\alpha=1.5$ ($\alpha>1$), obtained with $L=1000$, $m_0=0.9,$ and $\gamma=0.6$, averaged over $10^6$ realisations. The red dashed lines correspond to the analytical predictions in Eq.~\eqref{eq:8} and Eq.~\eqref{eq:10}, and the blue curves correspond to the numerical simulations.}
  \label{fig:1}
\end{figure}

The analytical predictions are in a very good agreement with the numerical simulations, as shown in Figure~\ref{fig:1}. Subfigure~\ref{fig:1a} corresponds to $\alpha<1$, where the system does not reach a steady state and the distribution at a finite time exhibits the expected double-peaked profile. Subfigure~\ref{fig:1b} corresponds to $\alpha>1$, where the system reaches a stationary distribution featuring similar divergences at $m=0$ and near $m=m_0$ as in the $\alpha<1$ case. The numerical curve again closely follows the theoretical forms. A broadening of the simulation peaks (especially near $m=0$) arises from finite size effects and the statistical fluctuations inherent in sampling extremely low-probability events. For $\alpha>1$, a larger number of realisations is required to capture rare magnetisation values, which is why in simulations we have used a high number of iterations to obtain a smooth stationary curve. Overall, the simulations validate the analytical structure of the distributions across both regimes.

\section{ Resetting in the 2D Ising model}
\label{sec:4}
The Ising model exhibits a phase transition in two dimensions as a function of temperature, characterized by a ferromagnetic state for temperatures below the critical temperature $T_C$ and a paramagnetic phase for $T > T_C$. The introduction of exponential resetting \cite{magoni2020ising} with a resetting rate $r$ generates a new pseudo-ferro phase, in addition to the ferro and para phases, where the behaviour is similar to the original system without reset. Here, we explore how these behaviours change when the inter-reset distribution follows a power law.
 \par
 In the 2D Ising model, Glauber dynamics is not exactly solvable as in 1D. For performing the calculations, we use the well-established late-time approximations \cite{stauffer1997relaxation}, as used in \cite{magoni2020ising}. In the absence of resetting, the time dependence of the magnetisation is known to follow: $(i)$ For $T > T_C$, at large times $m(t) \approx a_1 e^{-\lambda t}$ and $a_1\approx m_0$, $(ii)$ For $T < T_C$,  $m(t)\approx m_{eq}\pm a e^{-bt^{c}}$, where $a$, $b$ and $c$ are constants that can be numerically determined. $(iii)$ For $T = T_C$, at large times $m(t)\approx b_c t^{-\phi}$, where $\phi=\beta /\nu z$. Here, $\nu$ is the Ising critical exponent related to the correlation length, $\beta$ is the correlation length associated with the order parameter, and $z$ is the dynamical critical exponent associated with the Ising-Glauber dynamics at $T=T_C$. This approximation does not hold for smaller values of time. Therefore, for early times, guided by numerics, we approximate the magnetisation as $m(t)\approx m_0+At^\zeta$. The parameters $A$ and $\zeta$ can be fixed by matching to the known long-time behaviour.
\par
$(i)$ We first consider the case $T>T_C$, where the steady state of the system without resetting is the paramagnetic state $m=0$. Starting from the initial state with magnetisation $m_0$, the system is expected to decay towards equilibrium exponentially, and at large times $m(t) \approx a_1 e^{-\lambda t}$. We assume that this form is true at all times and take $a_1 = m_0$. The
calculations here are along the lines of those for one dimension. For $\alpha<1$, the probability density at time $t$ is
\begin{equation}
    P_r(m,t)=\frac{sin(\pi \alpha)}{\pi} \frac{1}{m}\left[ ln\left(\frac{m_0}{m}\right)\right]^{-\alpha}\left[t \lambda-ln\left(\frac{m_0}{m}\right)\right]^{\alpha-1}
    \label{eq:11}
\end{equation}

As in the 1D model, no steady state exists in this regime. The distribution diverges both at $m=0$ and initial magnetisation $m_0$, reflecting the dominance of very short and very long intervals between resets.
When $\alpha>1$, we have
\begin{equation}
    P_r^{stat}(m)=\left (\frac{\alpha-1}{\alpha} \right) \left(\frac{1}{\left[\tau_0 \lambda\right]^{1-\alpha}}\frac{1}{m}\left[ ln\left(\frac{m_0}{m}\right)\right]^{-\alpha}+\delta(m-m_{\tau_0})\right)
    \label{eq:12}
\end{equation}
The above expressions are similar to those seen in the one-dimensional case discussed above, and we refer to this phase with coexistence of near-zero and near-initial magnetisation as a QF state.
\begin{figure}[htbp]
  \centering
  \begin{subfigure}[b]{0.5\textwidth}
    \includegraphics[width=\textwidth]{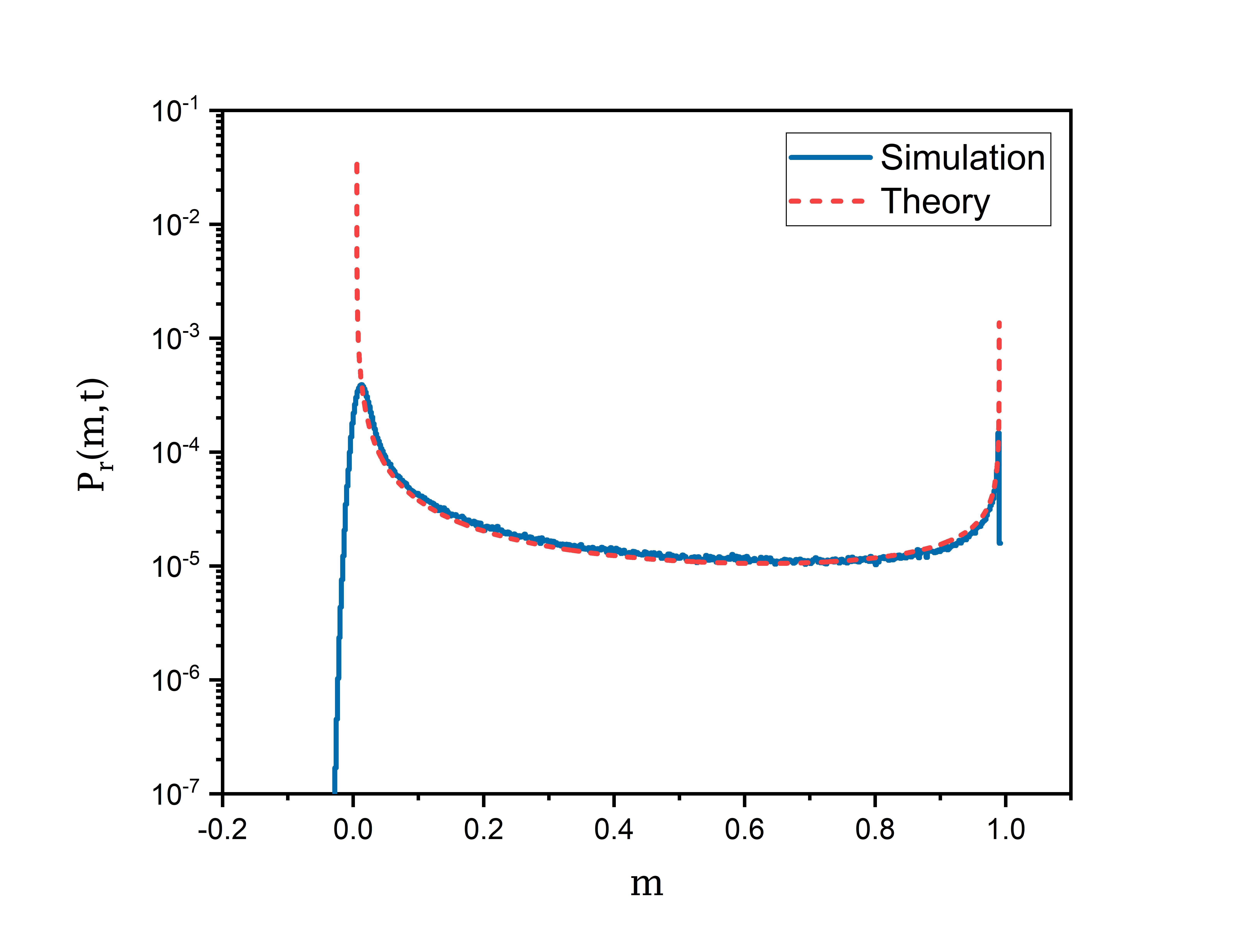}
    \caption{$\alpha<1$}
    \label{fig:2a}
  \end{subfigure}
  \hspace{-0.5cm}
  \begin{subfigure}[b]{0.5\textwidth}
    \includegraphics[width=\textwidth]{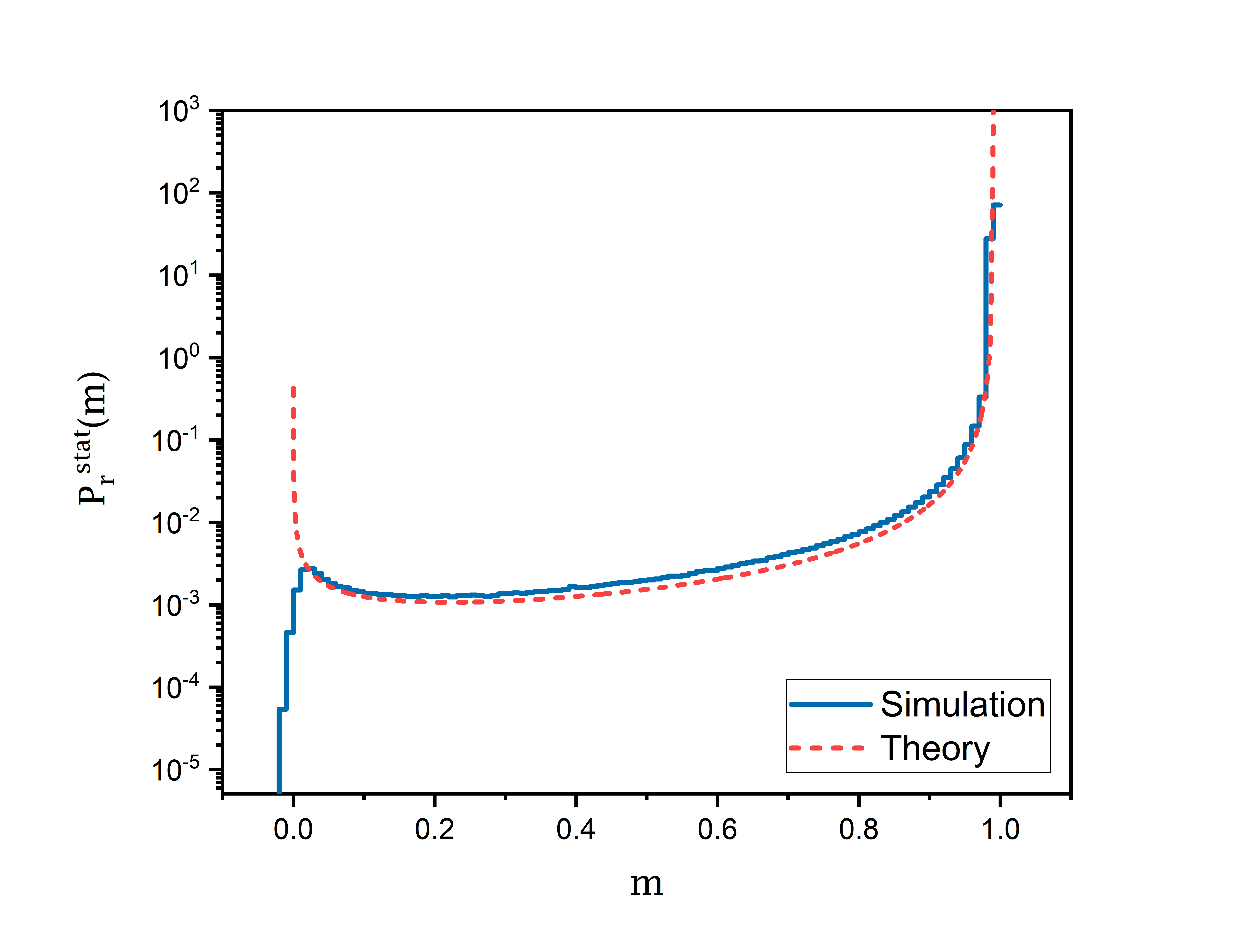}
    \caption{$\alpha>1$}
    \label{fig:2b}
  \end{subfigure}
  \caption{Plot of the PDF of the average magnetisation in the 2D Ising model of lattice size $256\times 256$ with power law resetting at a temperature $T>T_C$ ($T=3.5$) (a) Finite-time distribution at $t=40$ for $\alpha=0.5$ ($\alpha<1$) and $m_0=0.99$, averaged over $10^6$ independent realisations. (b) Stationary distribution for $\alpha=1.5$ ($\alpha>1$) and $m_0=0.99$, averaged over $2\times10^8$ realisations.}
  \label{fig:2}
\end{figure}
\par
To generate the theoretical curves shown in Figure~\ref{fig:2}, we first simulated the $2D$ Ising model without resetting for $T>T_C$ at $T=3.5$ on a $256\times256$ lattice starting from an initial magnetisation $m_0=0.99$. The relaxation of the magnetisation was fitted to the exponential form $m(t) = a_1 e^{-\lambda t}$, yielding $\lambda\approx0.113$, and we took $a_1=0.99$. These fitted parameters were then used in the analytical expressions for the power law resetting dynamics. Figure~\ref{fig:2} compares the theoretical predictions Eq.~\eqref{eq:11} and Eq.~\eqref{eq:12} with numerical simulations at $T=3.5$. For $\alpha<1$, the finite-time distribution at 
$t=40$ shows the expected divergences at $m=0$ and $m=m_0$. For $\alpha>1$, the system reaches a stationary distribution exhibiting the same divergences, corresponding to the QF state discussed earlier. In both cases, the agreement between theory and simulations is excellent. As before, we see the broadening of the $m=0$ peak.
\par
\begin{figure}[htbp]
  \centering
  % Figure 1
  \begin{subfigure}[b]{0.333\textwidth}
    \includegraphics[width=\linewidth]{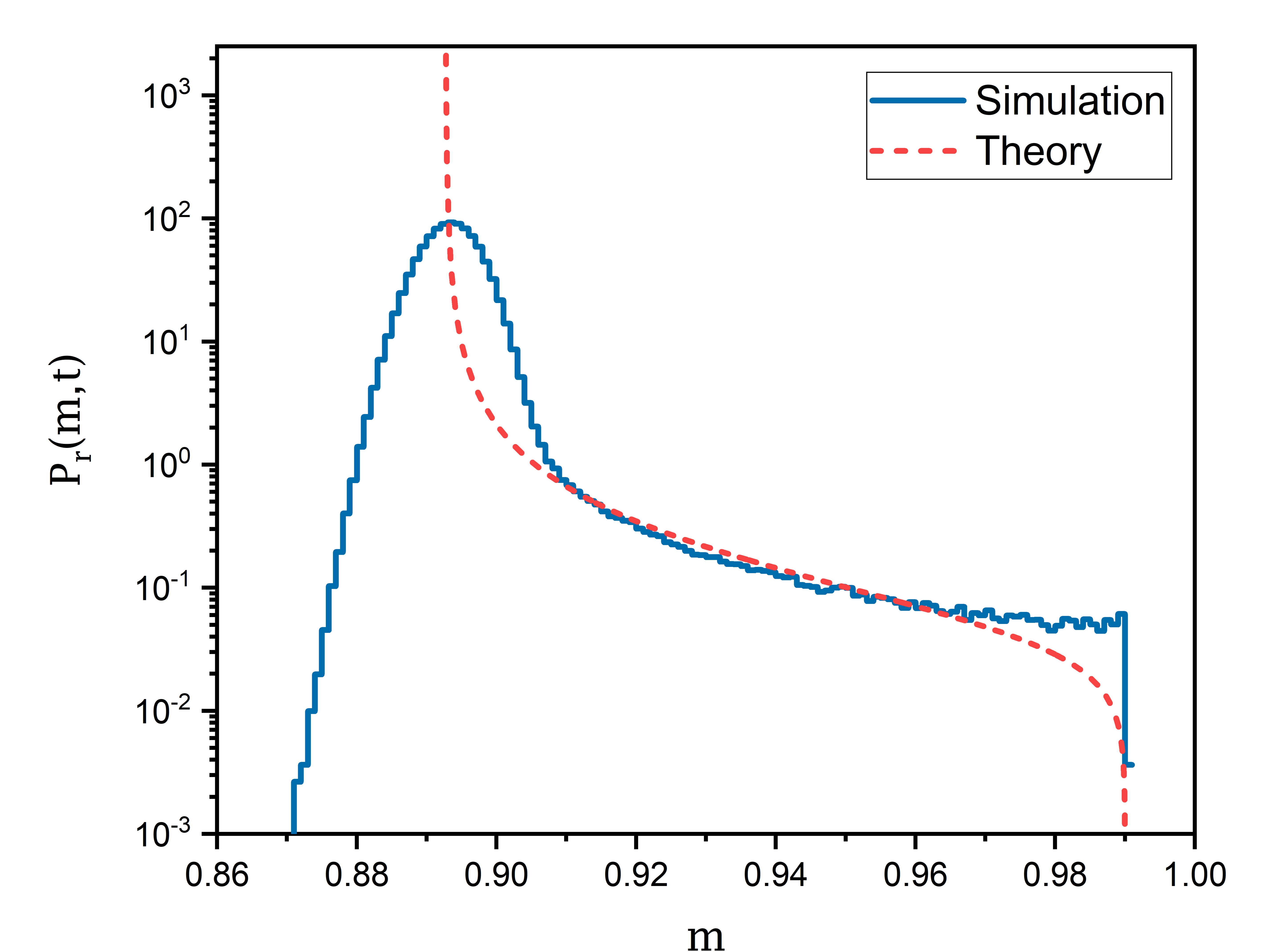}
    \caption{$\alpha<1$, $\alpha<\alpha^*$}
    \label{fig:3a}
  \end{subfigure}
  \hspace{-0.35cm}
  % Figure 2
  \begin{subfigure}[b]{0.333\textwidth}
    \includegraphics[width=\linewidth]{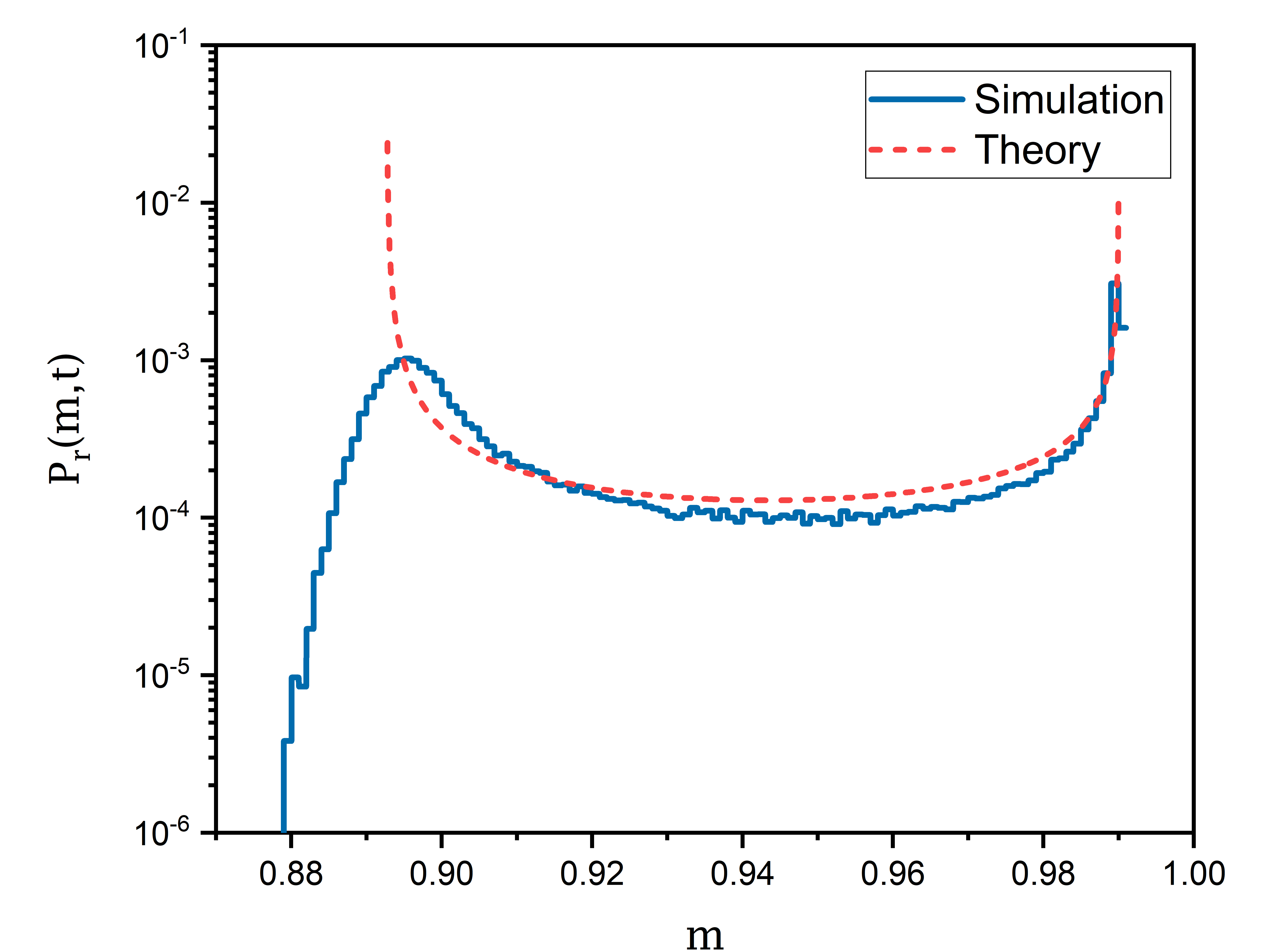}
    \caption{$\alpha<1$, $\alpha>\alpha^*$}
    \label{fig:3b}
  \end{subfigure}
  \hspace{-0.35cm}
  % Figure 3
  \begin{subfigure}[b]{0.333\textwidth}
    \includegraphics[width=\linewidth]{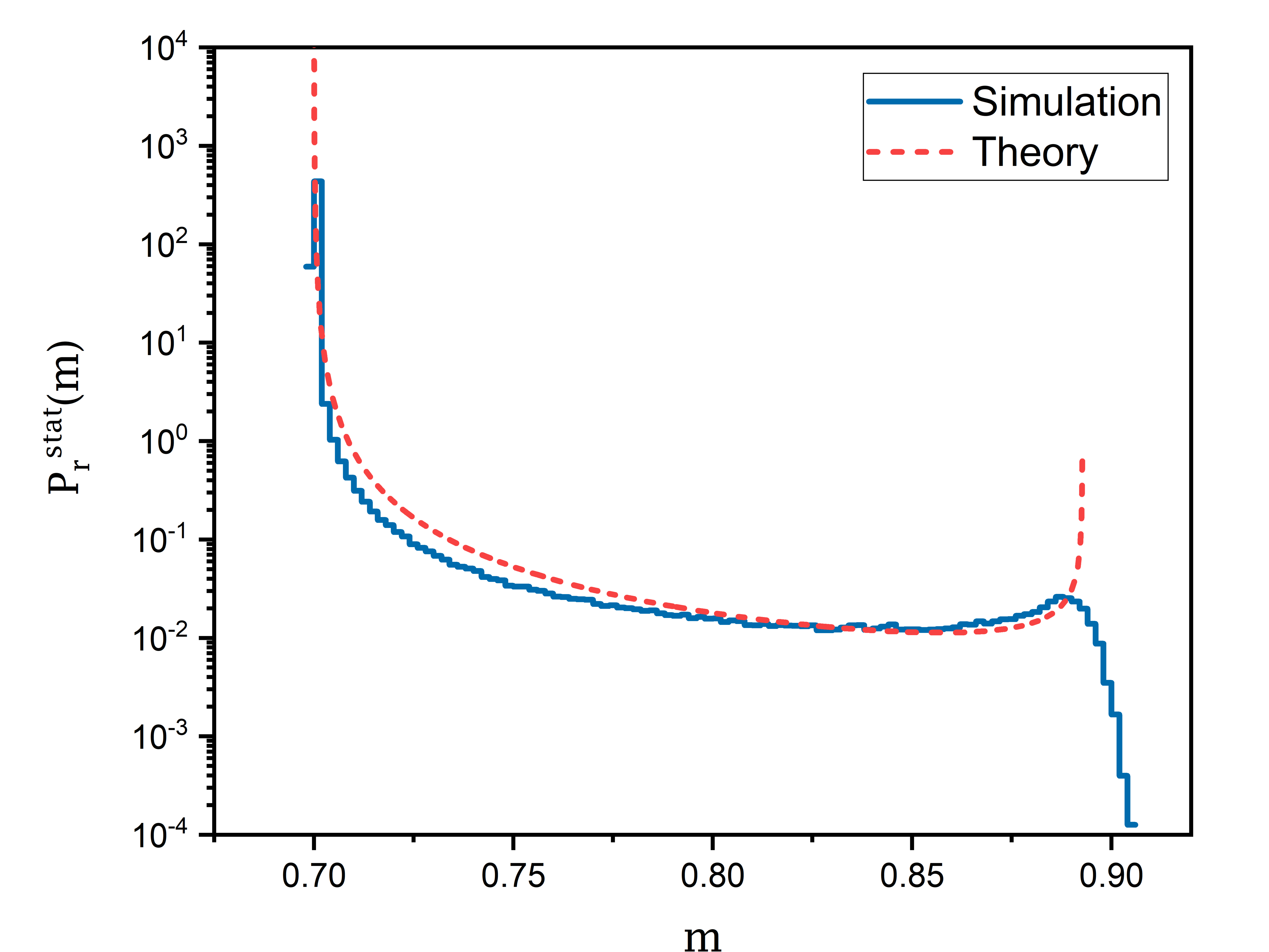}
    \caption{$\alpha>1$, $\alpha>\alpha^*$}
    \label{fig:3c}
  \end{subfigure}

  \caption{Probability density function (PDF) of the average magnetization in the two-dimensional Ising model with power law stochastic resetting for temperatures below the critical point (\(T < T_c\)). 
        System size \(L = 256\times256\). 
        (a) Distribution at \(t = 120\) for \( \alpha = 0.1 < \alpha^* \), \( T = 2.05 \), and \( m_0 = 0.99 \), averaged over \(3\times10^6\) realizations, with equilibrium magnetization \(m_{\mathrm{eq}} = 0.8927\). 
        (b) Distribution at \(t = 120\) for \( \alpha = 0.8 > \alpha^* \), \( m_0 = 0.99 \), and \( m_{\mathrm{eq}} = 0.8927\), averaged over \(10^5\) realizations. 
        (c) Steady-state distribution for \( \alpha = 1.5  \), obtained with \( m_0 = 0.7 \) and averaged over \(2\times10^7\) realizations.
    }
  \label{fig:3}
\end{figure}

$(ii)$ For $T<T_C$, if there is no resetting, the equilibrium state of the system is a ferromagnetic state with a finite magnetisation
$m_{eq}$. A system initially having a different magnetisation from $m_{eq}$ will approach this equilibrium value at
long times. The approach takes a stretched exponential form at large times $m(t)\approx m_{eq}\pm a e^{-bt^{c}}$. The values of $a,b,$
and $c$ ($0<c<1$) are not known analytically but can be determined from simulations. In the calculation below, we will
consider this stretched exponential form to be valid at all times (and therefore $a= (m_0\pm m_{eq})$, so that $m(t)=m_0$ at $t=0$).
When resetting is introduced, for $\alpha<1$, we get 
\begin{equation}
    P_r(m,t)=\frac{sin(\pi \alpha)}{\pi bc} \frac{1}{\pm(m-m_{eq})}\left[\frac{1}{b} ln\left(\frac{m_0-m_{eq}}{m-m_{eq}}\right)\right]^{\frac{1-c-\alpha}{c}}\left[t-\left(\frac{1}{b}ln\left(\frac{m_0-m_{eq}}{m-m_{eq}}\right)\right)^{\frac{1}{c}}\right]^{\alpha-1}
    \label{eq:13}
\end{equation}
There is a divergence at $m=m_{eq}$ as evident from the $ \frac{1}{m-m_{eq}}$ term (see figure~\ref{fig:3}). Another possibility
for a peak arises at $m=m_0$. This will happen when $1-\alpha -c<0$. Thus, we see a change in the behaviour in our
system as the value of $\alpha$ changes. The behaviour changes from a single peak at $m=m_{eq}$ as in subfigure~\ref{fig:3a} to a double peaked structure as seen in subfigure~\ref{fig:3b} as $\alpha$ crosses the value $\alpha^*=1-c$. When $\alpha>1$, in the limit of large time, we get the steady state expression,
\begin{equation}
    P_r^{stat}(m)=\left (\frac{\alpha-1}{\alpha} \right) \left( \frac{1}{bc \tau_0^{1-\alpha}}\frac{1}{\pm(m-m_{eq})}\left[\frac{1}{b} ln\left(\frac{m_0-m_{eq}}{m-m_{eq}}\right)\right]^{\frac{1-c-\alpha}{c}}+\delta(m-m_{\tau_0})\right)
    \label{eq:14}
\end{equation}
where we have used $a= (m_0\pm m_{eq})$. One sees that there is a divergence here at $m=m_{eq}$ and $m=m_0$ as can be seen in subfigure~\ref{fig:3c}. We thus have a distribution with a double peak again.
\par
The above results show that the system undergoes a change in behaviour at $\alpha=\alpha^*$ ($\alpha^*=1-c$). When $\alpha<\alpha^* $, the system exhibits a single peak at $m_{eq}$ and we refer to this state as a `single-peak ferro state' (SPF), whereas if $\alpha>\alpha^*$, the system shows two peaks at $m_0$ and $m_{eq}$ and the system is thus labeled as the `double-peak ferro state' (DPF). The DPF can be further split into two parts: for $\alpha<1$, there is no steady state, but for $\alpha>1$, we do have a steady state. The feature of dual peaks in the distribution of magnetisation persists for all values of $\alpha$

\par

Figure~\ref{fig:3} illustrates the behaviour of the probability density function of the average magnetisation in the two-dimensional Ising model with power law stochastic resetting for $T<T_C$. We first simulated the system without resetting at $T=2.05$ on a $256 \times 256$ lattice, for which the equilibrium magnetisation is $m_{eq}=0.8927$. Using an initial magnetisation $m_0=0.99>m_{eq}$, we fitted the relaxation curve to the stretched-exponential form $m(t)= m_{eq}+ a e^{-bt^{c}}$ and obtained $b=0.432$ and $c=0.591$, giving the threshold value $\alpha^* = 1-c = 0.409$. We used these values for the calculation of the PDF for the $\alpha<1$ case (Eq.~\eqref{eq:13}). A second fit was performed using a smaller initial magnetisation $m_0=0.7<m_{eq}$ (here $m(t)= m_{eq}- a e^{-bt^{c}}$), yielding $b=0.414$ and $c=0.718$; these parameters were used in the analysis of the $\alpha>1$ regime (Eq.~\eqref{eq:14}). Subfigure~\ref{fig:3a} shows the distribution at time $t=120$ for $\alpha=0.1<\alpha^*$, obtained using $m_0=0.99$ and averaged over $3\times10^6$ realisations; the distribution exhibits a single pronounced peak at $m_{eq}$. Subfigure~\ref{fig:3c} displays the distribution at the same time for $\alpha=0.8>\alpha^*$, again with $m_0=0.99$, averaged over $10^5$ realisations; here, a second peak emerges near $m_0$, signalling the onset of the double-peak ferro regime. Finally, subfigure~\ref{fig:3c} presents the steady-state distribution for $\alpha=1.5$, obtained using $m_0=0.7$ and averaged over $2\times10^7$ realisations, where both $m_0$ and $m_{eq}$ appear as stable long-time magnetisation values. The simulations validate the theoretical predictions, showing good correspondence throughout.

\par
$(iii)$ The system without reset undergoes a continuous phase transition at $T=T_C$ and a power law dependence is seen
for various quantities, with associated critical exponents. We are interested in the behaviour of magnetisation as a function of time, which in the limit of late times, shows the behaviour $m(t)\approx b_c t^{-\phi}$, with $\phi=\beta/\nu z$. For smaller times, we use the approximation, $m(t)\approx m_0+At^\zeta$. Once we have the values of $b_c$ and $\phi$, the parameters $A$ and $\zeta$ can be calculated by equating $m(t)$ and $P_r(m,t)$ at a small time $t'$ (here $t'$ is chosen as the earliest time at which the large-time approximation of $m(t)$ begins to accurately describe the numerical data). We use these approximate expressions in our calculations below.
With resetting and $\alpha<1$, we have
\begin{equation}
    P_r(m,t) =
\begin{cases} 
  \frac{sin(\pi \alpha)}{-\pi \zeta A} \left(\frac{\pm(m-m_0)}{A}\right)^{\frac{1-\zeta-\alpha}{\zeta}}\left[t-\left(\frac{\pm(m-m_0)}{A}\right)^{\frac{1}{\zeta}}\right]^{\alpha-1}, & \text{if } m_0\ge m\ge m(t')  \\[2mm]
  
  \frac{sin(\pi \alpha)}{\pi  \phi b_c} \left(\frac{m}{b_c}\right)^{\frac{\alpha-1-\phi}{\phi}}\left[t-\left(\frac{m}{b_c}\right)^{\frac{-1}{\phi}}\right]^{\alpha-1}, & \text{if } m<m(t') \\[1mm]
  \label{eq:15}
\end{cases}
\end{equation}
For magnetisation values close to the initial condition $m_0$, the distribution is governed by the early-time behaviour $m(t)\approx m_0+At^\zeta$, leading to a power-law dependence with exponent $(1-\zeta-\alpha)/\zeta$. For smaller magnetisation values, the system enters the late-time critical regime $m(t)\approx b_c t^{-\phi}$, which produces a different power law form with exponent $(\alpha-1-\phi)/\phi$. In both regions, one sees that the distribution exhibits a power law decay at large times, with the exponents determined jointly by the Ising dynamics ($\phi,\zeta$) and the resetting dynamics ($\alpha$).
For $\alpha>1$, we have the following expression in the limit of steady state
\begin{equation}
    [P_r^{stat}(m) =
\begin{cases} 
  \left (\frac{\alpha-1}{\alpha} \right) \left[ \frac{-1}{A\zeta \tau_0^{1-\alpha}}\left(\frac{\pm(m-m_0)}{A}\right)^{\frac{1-\zeta-\alpha}{\zeta}}+\delta(m-m_{\tau_0})\right], & \text{if } m_0\ge m\ge m(t')  \\[2mm]
  
   \left (\frac{\alpha-1}{\alpha} \right) \left[ \frac{1}{b_c \phi \tau_0^{1-\alpha}}\left(\frac{m}{b_c}\right)^{\frac{\alpha-1-\phi}{\phi}}+\delta(m-m_{\tau_0})\right], & \text{if } m<m(t') \\[1mm]
   \label{eq:16}
\end{cases}
\end{equation}
We see here an interesting crossover from a power law decay when $\alpha-\phi-1<0$, to a power law increase when
$\alpha-\phi-1>0$, with a flat distribution at $\alpha=\phi+1$. 
\begin{figure}[t]
  \centering
  \begin{subfigure}[b]{0.5\textwidth}
    \includegraphics[width=\textwidth]{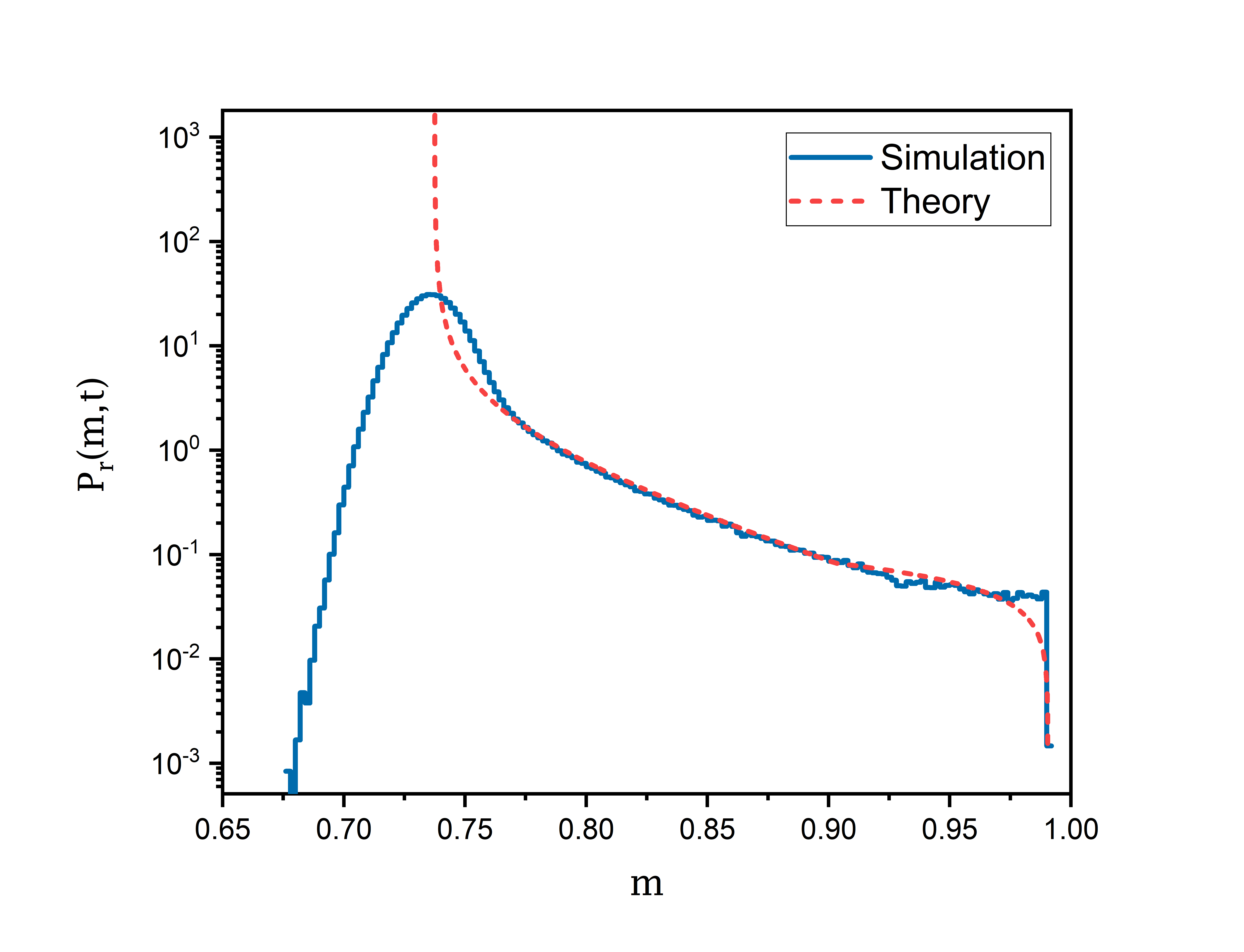}
    \caption{$\alpha<1$}
    \label{fig:4a}
  \end{subfigure}
  \hspace{-0.5cm}
  \begin{subfigure}[b]{0.5\textwidth}
    \includegraphics[width=\textwidth]{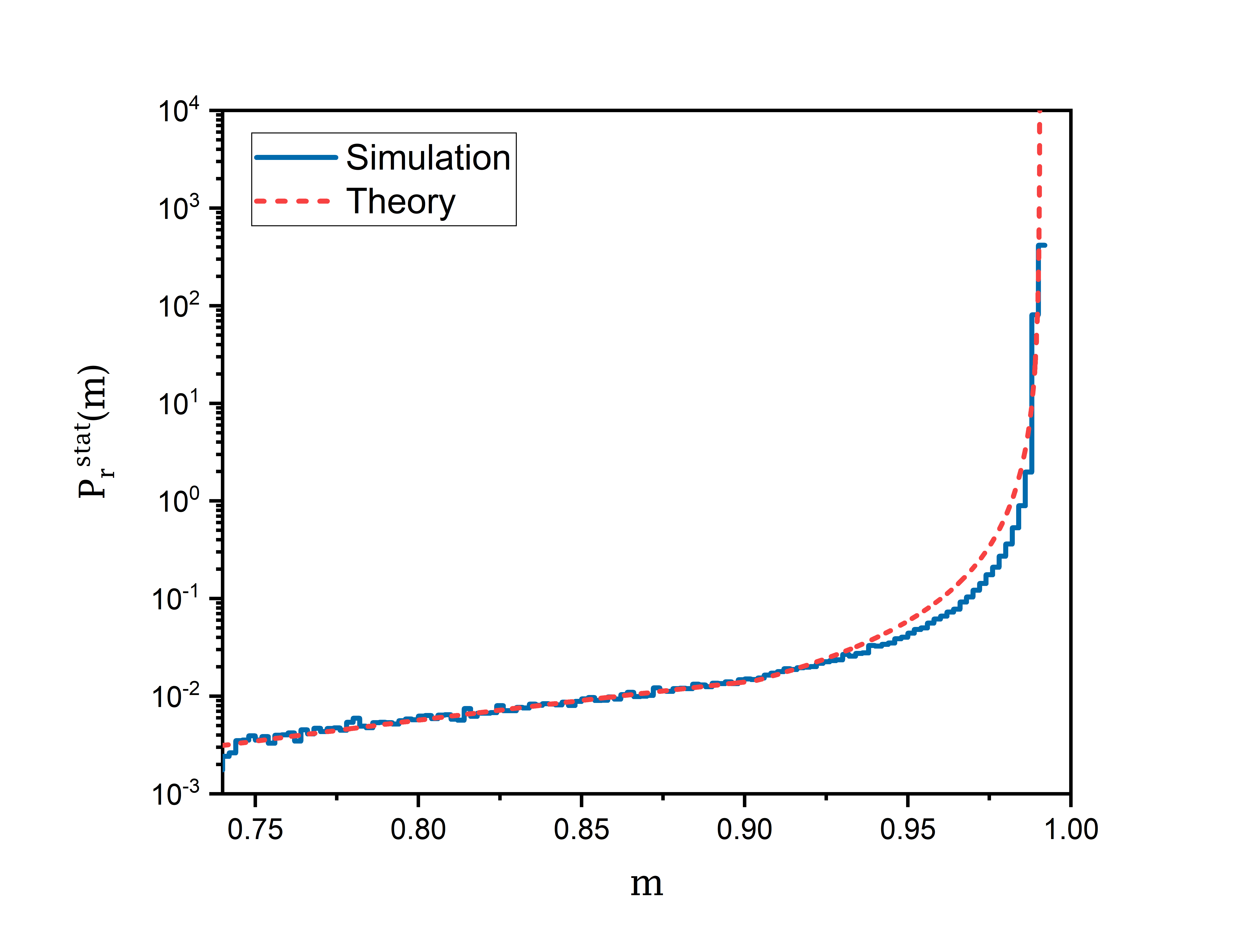}
    \caption{$\alpha>1$}
    \label{fig:4b}
  \end{subfigure}
  \caption{ Probability density function (PDF) of the average magnetization in the two-dimensional Ising model with power law stochastic resetting at the critical temperature \( T = T_c = 2.269 \). 
        System size \(L = 256\times256\). 
        (a) Distribution at \(t = 100\) for \( \alpha = 0.1 \), \( m_0 = 0.99 \), averaged over \(2.4\times10^6\) realizations. 
        (b) Time-independent distribution for \( \alpha = 1.5 \), obtained with \( m_0 = 0.99 \) and averaged over \(2\times10^7\) realizations.
    }
  \label{fig:4}
\end{figure}

To obtain the input parameters required for the analytical predictions, we first simulated the 2D Ising model without resetting at $T=T_C$ ($T_c = 2.269$ with $J = k_B = 1$) on a $256\times256$ lattice, starting from \(m_0 = 0.99\). The long-time relaxation of the magnetisation follows the critical decay form \(m(t) \approx b_c t^{-\phi}\), yielding \(b_c = 0.9624\) and \(\phi = 0.0578\). This approximation becomes accurate only beyond \(t' = 3\); for short times, data fits well to the \(m(t) \approx m_0 + A t^\zeta\). Matching the two regimes at \(t = t'\) gave \(A = -0.0452\) and \(\zeta = 0.5981\), which were subsequently used to construct the analytical PDFs. Subfigure~\ref{fig:4a} shows the magnetisation distribution at time \(t = 100\) for a heavy-tailed resetting distribution with exponent \(\alpha = 0.1\), averaged over \(2.4\times10^6\) realisations. Subfigure~\ref{fig:4b} presents the time-independent steady-state distribution obtained for \(\alpha = 1.5\), using the same initial condition \(m_0 = 0.99\) and averaging over \(2\times10^7\) realisations. These figures illustrate the close agreement between simulations and the analytically constructed PDFs.

\section{Conclusion}
\label{sec:5}
In summary, we have shown that introducing power law stochastic resetting to the Ising–Glauber dynamics leads to qualitatively new nonequilibrium behaviour that cannot arise either without resetting or under exponential-reset protocols. By analysing the relaxation forms of the magnetisation and incorporating them into the renewal framework, we obtained explicit expressions for the magnetisation distribution in both one and two dimensions. These theoretical predictions were validated through simulations, and together they reveal a rich structure of phases characterised by nontrivial divergences and the coexistence of peaks at the reset value and the equilibrium magnetisation.
\par
In the 2D Ising model, for $T>T_C$, the system exhibits a state with a persistent double-peaked magnetisation distribution for all values of $\alpha$. While the distribution is time-dependent for $\alpha<1$ and stationary for $\alpha>1$, this change does not affect the double-peak structure. We therefore classify this entire regime as a single phase, the quasi-ferro state (QF). This state is very different from the usual paramagnetic state expected in the absence of resetting, and also from the para and pseudo-ferro states seen in the presence of exponentially distributed inter-reset times. For $T<T_C$, varying $\alpha$ leads to a crossover at $\alpha=\alpha^*$ from a single-peaked (SPF) state with a magnetisation peak at $m_{eq}$, to a double-peaked (DPF) state with two peaks in the magnetisation distribution at $m=m_{eq}$ and $m=m_0$. Notice again that this behaviour is very different from the simple ferromagnetic behaviour seen in the Ising model without reset and also from the ferro state seen in the previous work with exponential resetting, where there is no such crossover. While the existence of a steady state emerges only for $\alpha>1$, the structure of the magnetisation distribution is governed by $\alpha^{*}$, rather than by steady-state considerations alone, and therefore we continue with the DPF label for all $\alpha>\alpha^*$. At the critical temperature $T=T_C$, the magnetisation distribution displays distinct power law regimes arising from early-time and late-time critical relaxation. For $\alpha<1$, the distribution remains time-dependent, while for $\alpha>1$ it reaches a stationary form. In the stationary regime, the exponent $\alpha-\phi-1$ controls a crossover from a decaying to a growing power law magnetisation distribution, with a flat distribution at $\alpha=\phi+1$. For the 1D Ising model, where the system remains paramagnetic at all finite temperatures, power-law resetting similarly generates a QF, with a steady-state distribution emerging for $\alpha>1$.

\par
Thus, we see that the behaviour reported here is remarkably rich and novel. We expect that the general features seen here will be relevant beyond the Ising model. Our results highlight the sensitivity of interacting many-body systems to the statistics of resetting events and suggest several natural extensions. These include investigating other spin systems or higher dimensions, and analysing spatial correlations and domain growth under power law resets. More generally, our findings suggest that non-Poissonian resetting can be used as a general tool in statistical physics to engineer and control nonequilibrium states.
\bibliographystyle{unsrt}
\bibliography{ref}

\end{document}